# Carbon Nanocone: A Promising Thermal Rectifier

Nuo Yang[1], Gang Zhang[2, a)] and Baowen Li[3,1, b)]

[1]*Department of Physics and Centre for Computational Science and Engineering,*

*National University of Singapore, 117542 Singapore*

[2]*Institute of Microelectronics, 11 Science Park Road, Singapore Science Park II,*

*Singapore 117685, Singapore*

[3]*NUS Graduate School for Integrative Sciences and Engineering, 117597 Singapore*

## ABSTRACT

With molecular dynamics simulations, we demonstrate that the carbon nanocone is an excellent thermal rectifier. Obvious thermal rectification ratio in large temperature range, from 200K to 400K, has been observed. Furthermore, the rectification of nanocone does not depend on the length very sensitively, which is in stark contrast with the nanotube thermal rectifier in which the rectification decreases dramatically as the length increases. In nanocone, the heat flux is controlled by match/mismatch of the phonon power spectra. Our work demonstrates that carbon nanocone is a promising practical phononic device.

a) Electronic mail: zhangg@ime.a-star.edu.sg

b) Electronic mail: phylibw@nus.edu.sg

Traditionally, information is carried and processed by electrons and photons. Very recently, phononic (thermal) devices have been brought forward theoretically, in which phonon– a heat pulse through lattice**,** is used to carry and process information, [1-7] which has added a dimension to information science and technology in addition to electronics and photonics. Elementary phononic devices such as diode, transistor and logic devices have been proposed. In addition to information processing, the thermal devices might also have broad applications for heat control/management in the future. It has attracted an increasing interest both from fundamental research as well as applied research in recently years. Similar to electronic counterparts, the thermal rectifier also plays a vital role in phononics circuit. [7] Some theoretical models [1-4] have been proposed for thermal rectifiers. However, we are far away from real devices because of many approximations and assumptions used in the models, and the low rectification efficiency. The solid-state thermal rectifier, carbon nanotubes with non-uniform axial mass distribution, has been reported experimentally by C. W. Chang *et al.* [8]. Wu and Li have also demonstrated the rectification in carbon nanotube *(n, 0)/(2n, 0)* intramolecular junctions by using molecular dynamics (MD) method. [9] However, the rectifications in both graded massed and graded structural carbon nanotube junctions are much lower than that of electronic counterparts. Moreover, in all theoretical models so far, the rectification efficiency decreases quickly as the structure length increases. This largely limits the application of the nanotube thermal rectifier. In the current paper, we report one nanostructure – carbon nanocone – as a potential candidate for realistic thermal rectifier. The large structural asymmetry in carbon nanocone (as shown in figure 1) enhances the rectification efficiency remarkably.

Carbon nanocone has a high asymmetric geometry. [10] The cone is entirely characterized by its cone angle. When one pentagon is introduced into a hexagonal carbon network, a 60º disclination defect is formed, leading to the formation of a nanocone with cone angle of 113º. In this work, we focus on the cone with the cone angle of 113$^o$, which is the largest angle observed experimentally [10, 11] and theoretically. [12]

In our simulations, classical non-equilibrium molecular dynamics method is adopted. The potential energy for carbon nanocone is a Morse bond and a harmonic cosine angle for bonding interaction, which include both two-body and three-body potential terms:

$$U\left(r_{ij},\theta_{ijk}\right) = K_{Cr}\left(e^{-\gamma\left(r_{ij}-r_C\right)}-1\right)^2 + K_{C\theta}\left(\cos\theta_{ijk}-\cos\theta_C\right)^2/2 \qquad (1)$$

where $\theta_{ijk}$ represent bending angles and $r_{ij}$ represent distances between bonded atoms. $K_{Cr}$ =478.9 $kJmol^{-1}\AA^2$, $K_{C\theta}$ =562.2 $kJmol^{-1}$ are the force constants of the stretch and bend potentials, respectively. The equilibrium carbon-carbon bond length, $r_C$, is 1.418 Å and the bond angle, $\theta_C$ =120$^o$, which are the corresponding reference geometry parameters for graphene. [13] The force field potential used in the current manuscript was developed by Y. Quo, N. Karasawa and W. A. Goddard III by fitting experimental lattice parameters, elastic constants and phonon frequencies for graphite to interpret various properties for fullerene (including vibrational spectroscopy, crystal structure analysis). [14] The accuracy of Morse potential is demonstrated as the calculated thermal conductivity of (10, 10) single wall carbon nanotube with 40 nm in length is 315 W/mK with Morse potential, which is very close to values ~350 W/mK [15], ~310 W/mK [16], and ~215 W/mK [17] reported by others.

First, the heat flux in nanocone with forty layers (N=40) is explored, whose length along the side face is around 5.6 nm. Here, fixed boundary conditions are used, that the atoms in the two end layers are frozen. In order to establish a temperature gradient along the longitudinal direction, the nanocone is coupled with Nosé-Hoover heat bathes [18, 19] at the *second and N-1*'th layers, with temperatures are $T_{top}$ and $T_{bottom}$ respectively. Besides Nosé-Hoover heat bath, other methods can also be used to produce a temperature gradient and calculate the heat flux, such as the method developed by Müller-Plathe in which a cold and a hot region of the system are created by switching the velocity of the hottest atom in the cold region with the velocity of the coldest atom in the hot region. Nevertheless, it has been demonstrated that the thermal conductivity and heat flux calculated by MD do not depend on the details of the heat bath used. [4, 15-17, 20] For instance, the thermal conductivities of silicon nanowires calculated with Nosé-Hoover heat bath agree very well with those calculated with Langevin heat bath. [20] In this paper, all the calculations are done with the same Nosé-Hoover heat bath parameter (thermostat response time) on the two ends.

Then the velocity Verlet algorithm is used to integrate the differential equations of motions. In general, the temperature, $T_{MD}$, is calculated from the kinetic energy of atoms according to the Boltzmann distribution:

$$\langle E \rangle = \sum_{1}^{N} \frac{1}{2} m v_i^2 = \frac{3}{2} N k_B T_{MD}, \qquad (2)$$

where $\langle E \rangle$ is the mean kinetic energy, $v_i$ is the velocity of atom, $m$ is the atomic mass, $N$ is the number of particles in the system, and $k_B$ is the Boltzmann constant. It is worth pointing out that this equation is valid only at very high temperature (**$T$>>$T_D$**, **$T_D$** is the Debye temperature). When the system average temperature is lower than the Debye temperature, it is necessary to apply a quantum correction to the MD calculated temperature. The difference between the MD calculated temperature and quantum corrected temperature depends on the Debye temperature. However, the Debye temperature of carbon nanocone is not known. Although many theoretical studies have been conducted on the Debye temperature of carbon nanotube, the results have been controversial. For instance, 473 K [21], 475 K [22], 580 K [23], and 1000 K [24] are reported in different literatures. The concerns of the current paper are the heat flux and rectification effects (a related change of heat flux) which do not depend on the accurate value of temperature-but the temperature difference. So, we don't do the quantum correction to $T_{MD}$ in this work.

The heat flux $J_l$ along the nanocone is defined as the energy transported along the surface in unit time. The expression of flux is

$$J_l(t) = \sum_i v_{i,l} \varepsilon_i + \frac{1}{2} \sum_{ij} r_{ij,l} \left( \overrightarrow{F}_{ij} \cdot \overrightarrow{v}_i \right) + \sum_{ijk} r_{ij,l} \left( \overrightarrow{F}_j(ijk) \cdot \overrightarrow{v}_j \right) \tag{3}$$

where $\varepsilon_i$ is local site energy, $\overrightarrow{F}_{ij}$ is two-body force, and $\overrightarrow{F}_j(ijk)$ is three-body force (details can be seen in [25]).

Simulations are performed long enough such that the system reaches a stationary state when the local heat flux reaches a constant along the nanocone. All results given in this work are obtained by averaging about $4 \times 10^6$ time steps. The time step is set as $0.4$ fs. The total thermal conductivity contains contributions from electrons and phonons. However, phonons dominate the heat transport in carbon nanocone because it is a semiconductor. [12]

In this paper, we set the temperature of top as $T_{top} = T_0(1-\Delta)$ and that of bottom as $T_{bottom} = T_0(1+\Delta)$, where $T_0$ is the average temperature, and $\Delta$ is the normalized temperature difference between the two ends. Therefore, the bottom of nanocone is at a higher temperature when $\Delta > 0$, and the top has a higher temperature when $\Delta < 0$. The heat flux $J$ versus temperature difference $\Delta$ (corresponds to the "I-V" curve of electric rectifier) is shown in figure 2a. When high temperature bath contacts with the bottom end ($\Delta > 0$), the heat flux ($J$) increases steeply with $\Delta$; while in the region $\Delta < 0$, the heat flux is much smaller and changes a little with $\Delta$. That is, the nanocone behaves as a "good" thermal conductor under positive "thermal bias" and as a "poor" thermal conductor under negative "thermal bias". It suggests that the heat flux runs preferentially along the direction of decreasing diameter. To describe quantitatively the rectifier efficiency we introduce the *thermal rectification*,

$$\text{R} \equiv \frac{(J_+ - J_-)}{J_-} \times 100\%$$

where $J_+$ is the heat current from bottom to top corresponds to $\Delta > 0$ and $J_-$ is the heat current from top to bottom corresponds to $\Delta < 0$. Figure 2b shows the rectifications with different $\Delta$. The increase of $\Delta$ results in the increase of the rectification ratio akin to the characteristic in electric rectifier.

In the solid-state thermal rectifier experiment, [8] carbon nanotubes are gradually deposited on the surface with heavy molecules along the length of the nanotube to establish asymmetric mass distribution. To increase the graded effects in carbon nanocone, we also studied the nanocone structure with the graded mass distribution. The mass of the atoms of the *ith* layer are set as $M_i = [1 + 3(i-1)/(N-1)]M_{C12}$, that is, the top atoms have minimum mass ($M_{C12}$) and the bottom atoms have maximum mass ($4M_{C12}$), where $M_{C12}$ is the mass of $^{12}$C atom. In Ref. 8, the atomic mass ratio is about 5 and the room temperature rectification is only 2% with $|\Delta| \approx 0.05$. However, the rectification ratio is 10% for uniform massed nanocone and 12% for graded massed nanocone with the same $\Delta$ value. With graded mass distribution, the rectification ratio increases with 2%, which agrees well with the experimental results of Ref. 8. At small temperature difference range, the geometric impact is more effective than the mass impact. And the rectification ratio of carbon nanocone is also much higher than that of carbon nanotube *(n, 0)/(2n, 0)* intramolecular junction. For example, at room temperature and with $|\Delta|=0.5$, the rectification of carbon nanocone is 96%, while that of carbon nanotube intramolecular junction is only about 15% (see figure 3 in Ref. 9). Our results demonstrate that carbon nanocone rectifier has obvious advantage over carbon nanotube based thermal rectifiers [8,9] in higher rectification ratio. And no additional mass loading process is required in the fabrication of carbon nanocone thermal rectifier.

The rectification versus temperature $T_0$ is shown in figure 3a. As the temperature increases, the heat fluxes increase, and the increase of $J_-$ is higher than the increase of $J_+$ (see figure 3b), which leads to the reduction of the rectification ratio. At low temperature ($T_{MD}$ from 200K to 300K), the small increase of temperature induces large reduction in rectification ratio. Contrast to the high dependence at the low temperature range, the rectification versus temperature curve is close to flat around the room temperature ($T_{MD}$ from 300K to 400K). It demonstrates carbon nanocone has significant rectification effect in a wide temperature range (about 100% at $T_{MD}$=300K, and 80% at $T_{MD}$=400K). Here the temperature ($T_{MD}$) range is from 200 K to 400K. Based on the results of quantum modification to $T_{MD}$ in Ref [21], the results presented here correspond to occur in real temperature range of 120 K to 360 K, are observable experimentally.

The results shown above are based on a cone structure with $N$=40 layers. In the following, we explore the impact of nanocone length on the rectification, with the longest nanocone side face length is around 14 nm ($N$=100). The nanocone length can be controlled experimentally in a narrow range, typically around 20 nm. [10] The 14 nm cone we explored here is in the range of the experimental observation. Figure 3c shows that the rectification is independent on the length. As shown in figure 3d, the heat fluxes are independent on the size, which means that energy transports ballistically in the system. In the molecular junction typed carbon nanotube thermal rectifier [9], the rectification decreases quickly when device length is increased, because the role of the interface is suppressed in large system. However, in the nanocone rectifier proposed here, high rectification can be achieved in the practical nanocone length scale, adds the feasibility of constructing thermal rectifier with carbon nanocones.

To understand the underlying mechanism of the rectification phenomenon, we calculated the power spectra. The power spectrum is a good way to explain the heat rectification effect both qualitatively and quantitatively. [26] The power spectrum is calculated from the Fourier transform of the velocity, and can be used in anharmonic system. We show the power spectra of atoms close to the top and the bottom (4th and 37th layers, the heat baths are contacted at 2nd layer and 39th layer, respectively) in figure 4. The frequency range of the power spectrum is from 0 to 60 THz, which is consistent with the reported vibration spectra [15, 27- 29] of carbon nanotubes. There is significant difference between nano and bulk material in the power spectra. In bulk materials, the optical modes contribute little to heat flux. However, in nano scale system such as carbon nanotube, anomalous thermal conductivity was observed which did not obey the Fourier law. [30] In the non-Fourier system with finite length, optical phonon (high frequency) also plays a major role to the heat flux. [27] As shown in figure 4a, where the bottom end is at high temperature ($\Delta > 0$), the power spectra of the top and bottom atomic layers overlap perfectly in a large frequency range, which means the phonon can easily go through the nanocone along the direction of temperature gradient, and correspond to large $J_+$. On the contrary, when $\Delta < 0$ (figure 4b), there is an obvious mismatch in the power spectra, both in low and high frequency band. The large mismatch in the spectra shows the weak correlation between the two ends. As a result, the phonon is difficult to go through the structure and thus leads to very small $J_-$. The match/mismatch of the power spectra between the bottom and top atomic layers controls the heat current and results in the rectification phenomenon. In order to quantify the above power spectrum analysis, the overlaps ($S$) of the power spectra of the two layers are calculated as [26]:

$$S_\pm = \frac{\int P_b(\omega) P_t(\omega) d\omega}{\int P_b(\omega) d\omega \int P_t(\omega) d\omega},$$

Here $S_\pm$ corresponds to the case of $\Delta > 0$ and $\Delta < 0$, respectively. And $P_t(\omega)$ is the power spectrum of the 4th (top) layer and $P_b(\omega)$ is the power spectrum of the 37th (bottom) layer. From figure 4, a $S_+ / S_-$ ratio of 1.2 is observed, corresponds to larger power spectra overlap when $\Delta > 0$. This picture indeed illustrates that the heat flux and rectification phenomena correlate strongly with the overlap of the power spectra of the two layers.

In conclusion, we have demonstrated an excellent thermal rectifier from carbon nanocone. Obvious thermal rectification ratio in a large temperature range, from 200K to 400K, has been observed. It shows that the graded geometric asymmetry is of remarkable benefit to improve the rectification ratio. Compared to the quick decrease of rectification ratio with device length in carbon nanotube junction rectifier, the weak dependence of rectification ratio on nanocone length is of particular importance for phononics application. In nanocone, the heat flux is controlled by match/mismatch of the power spectra between the top and bottom atomic layers. In contrary to the previous studies of thermal rectifiers which usually have three [1] or two coupled segments [2, 3], the nanocone can control the heat current by itself. More importantly, in carbon nanocone we don't need any on-site potential required in previous theoretical models, [1-3] which is hard to control experimentally, and moreover, the onsite (substrate) potential generally reduce the heat current significantly. These advantages of carbon nanocone thermal rectifier raise the exciting prospect that the practical phononics device can be realized with a full-carbon system.

This work is supported in part by an ARF grant, R-144-000-203-112, from the Ministry of Education of the Republic of Singapore, and Grant R-144-000-222-646 from National University of Singapore.

**Figure Captions**

Figure 1. Schematic picture of the carbon nanocone. $d$ is the distance from the atom to the center point in cone sheet. We define the atoms in the $i$th layer as the atoms whose $d \subset \left( (i-1)r_C, ir_C \right]$. The length along the side face of nanocone is $L = N \cdot r_C$ where N is the number of layers and $r_C$ is the equilibrium carbon-carbon bond length.

Figure 2. (a) Heat flux J versus $\Delta$. (b) Rectifications versus $\Delta$ for the carbon nanocone. Here N=40, and $T_0$=300 K.

Figure 3 (Color on-line) (a) and (b) Rectifications and heat fluxes of carbon nanocone at different temperature, $T_0$, with fixed $N = 40$ and $|\Delta| = 0.5$. (c) and (d) Rectifications and heat fluxes for different length of carbon nanocone, with fixed $T_0 = 300K$, and $|\Delta| = 0.5$. Here $N$ changes from 20 to 100.

Figure 4 (Color on-line) Power spectra of atoms at the 4th (close to the top, solid lines) and 37th (close to the bottom, dashed lines) atomic layers. The heat baths are contacted with 2nd and 39th layers, respectively. (a) $T_0$=300K, $\Delta$=0.5; (b) $T_0$=300K, $\Delta$=-0.5. It corresponds to big heat flux when $\Delta$>0, and corresponds to small heat flux with $\Delta$<0.

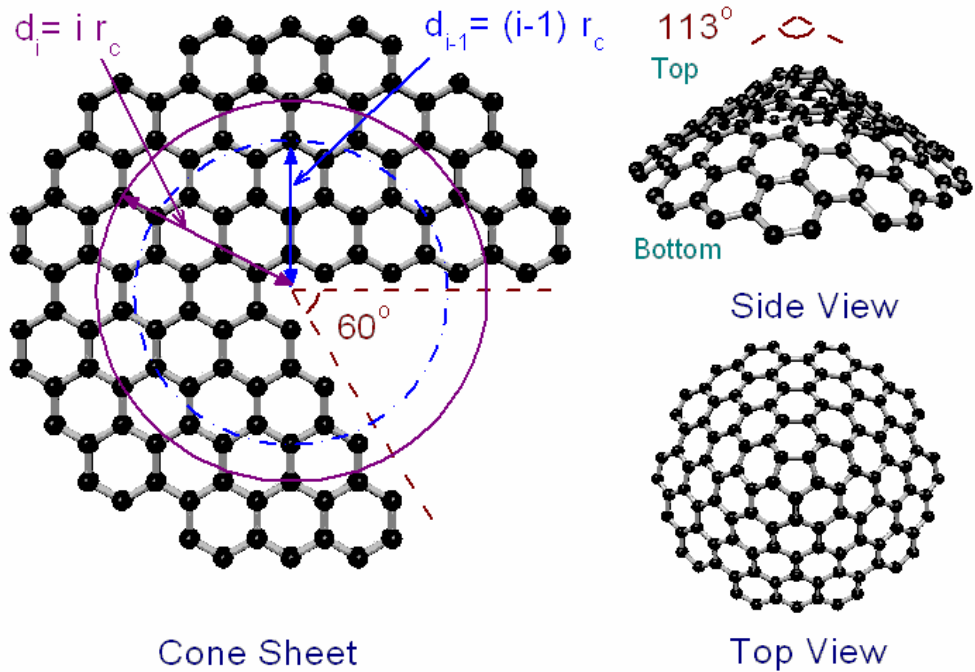

Figure 5. Schematic picture of the carbon nanocone. $d$ is the distance from the atom to the center point in cone sheet. We define the atoms in the $i$th layer as the atoms whose $d \subset ((i-1)r_C, ir_C]$. The length along the side face of nanocone is $L = N \cdot r_C$ where N is the number of layers and $r_C$ is the equilibrium carbon-carbon bond length.

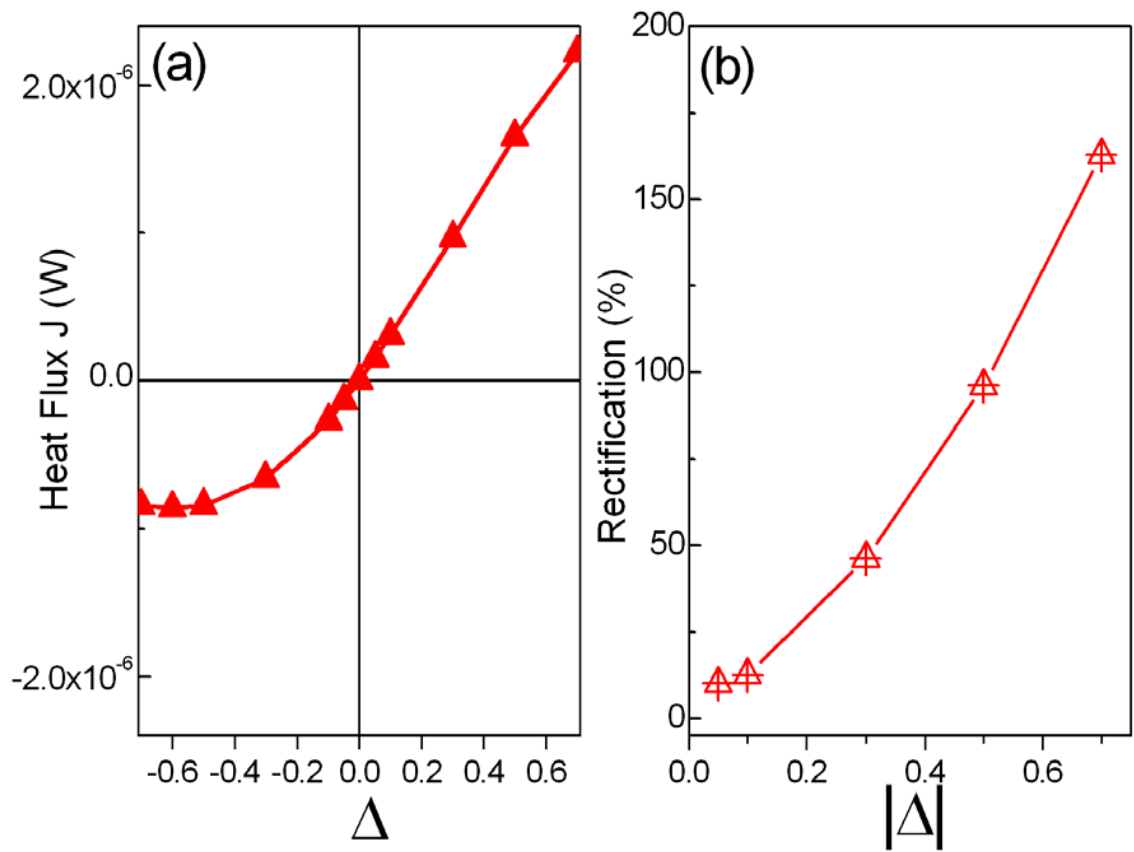

Figure 6. (a) Heat flux J versus $\Delta$. (b) Rectifications versus $\Delta$ for the carbon nanocone.

Here N=40, and $T_0$=300 K.

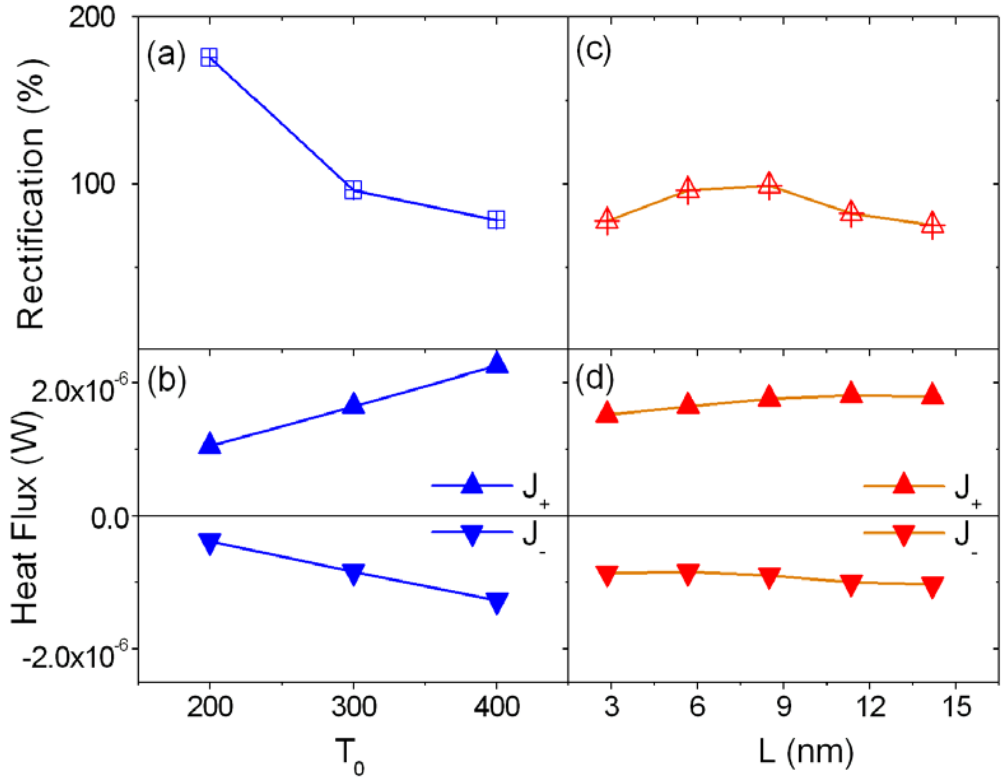

Figure 7 (Color on-line) (a) and (b) Rectifications and heat fluxes of carbon nanocone at different temperature, $T_0$, with fixed $N = 40$ and $|\Delta| = 0.5$. (c) and (d) Rectifications and heat fluxes for different length of carbon nanocone, with fixed $T_0 = 300K$, and $|\Delta| = 0.5$. Here $N$ changes from 20 to 100.

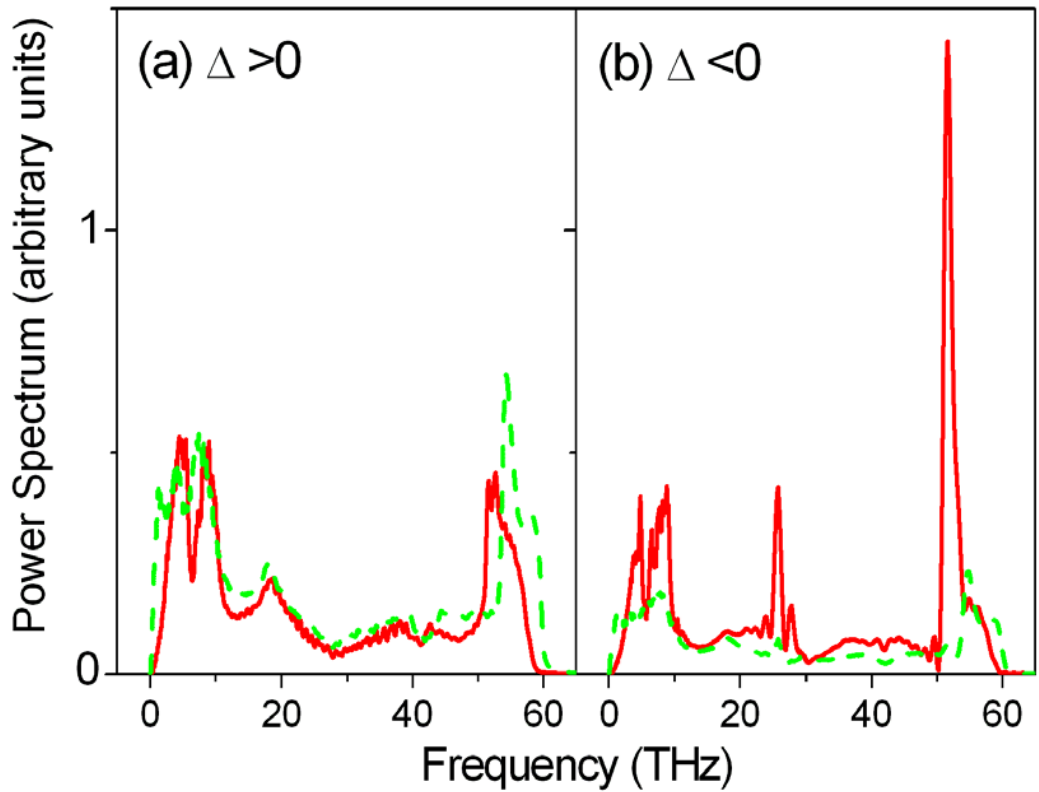

Figure 8 (Color on-line) Power spectra of atoms at the 4th (close to the top, solid lines) and 37th (close to the bottom, dashed lines) atomic layers. The heat baths are contacted with 2nd and 39th layers, respectively. (a) $T_0$=300K, $\Delta$=0.5; (b) $T_0$=300K, $\Delta$=-0.5. It corresponds to big heat flux when $\Delta$>0, and corresponds to small heat flux with $\Delta$<0.